\documentclass[a4paper,11pt]{article}

\usepackage{jheppub} 

\usepackage[T1]{fontenc} 

\usepackage{amssymb}
\usepackage{rotating}
\usepackage{adjustbox}
\usepackage{tabularx}
\usepackage{longtable}
\usepackage{lscape}
\usepackage{rotating}
\usepackage{supertabular,booktabs}
\usepackage{color}
\usepackage{amsmath}
\usepackage{wrapfig}
\usepackage{makecell}
\usepackage{subcaption}
\title{\boldmath Low Radioactive Material Screening and Background Control for the PandaX-4T Experiment}
\collaborationImg{
\begin{figure}[!htbp]
    \includegraphics[width=0.3\textwidth]{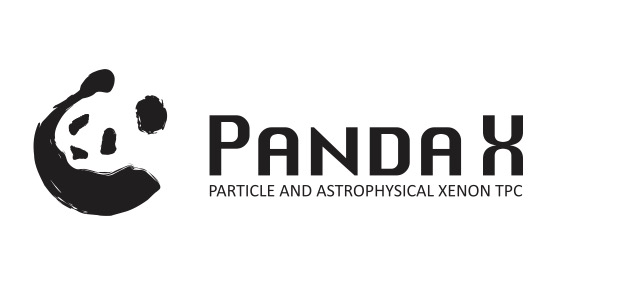}
\end{figure}
}
\author[a]{Zhicheng Qian}
\author[a]{Lin Si}
\author[a]{Abdusalam Abdukerim}
\author[a]{Zihao Bo}
\author[a]{Wei Chen}
\author[a,b]{Xun Chen}
\author[c]{Yunhua Chen}
\author[d]{Chen Cheng}
\author[e,f]{Yunshan Cheng}
\author[g]{Xiangyi Cui}
\author[h]{Yingjie Fan}
\author[i]{Deqing Fang}
\author[i]{Changbo Fu}
\author[j]{Mengting Fu}
\author[k,l,s]{Lisheng Geng}
\author[a]{Karl Giboni}
\author[a]{Linhui Gu}
\author[c]{Xuyuan Guo}
\author[a]{Ke Han}
\author[a]{Changda He}
\author[c]{Jinrong He}
\author[a]{Di Huang}
\author[m]{Yanlin Huang}
\author[a]{Zhou Huang}
\author[b]{Ruquan Hou}
\author[n]{Xiangdong Ji}
\author[o]{Yonglin Ju}
\author[a]{Chenxiang Li}
\author[c]{Mingchuan Li}
\author[o]{Shu Li}
\author[g]{Shuaijie Li}
\author[p,q]{Qing Lin}
\author[a,b,g]{Jianglai Liu}
\author[e,f]{Xiaoying Lu}
\author[j]{Lingyin Luo}
\author[a]{Wenbo Ma}
\author[i]{Yugang Ma}
\author[j]{Yajun Mao}
\author[a,b]{Yue Meng}
\author[a]{Xuyang Ning}
\author[c]{Ningchun Qi}
\author[e,f]{Xiangxiang Ren}
\author[e,f]{Nasir Shaheed}
\author[c]{Changsong Shang}
\author[k]{Guofang Shen}
\author[c]{Wenliang Sun}
\author[n]{Andi Tan}
\author[a,b]{Yi Tao}
\author[e,f]{Anqing Wang}
\author[e,f]{Meng Wang}
\author[i]{Qiuhong Wang}
\author[a,r]{Shaobo Wang}
\author[j]{Siguang Wang}
\author[d]{Wei Wang}
\author[o]{Xiuli Wang}
\author[a,b,g]{Zhou Wang}
\author[d]{Mengmeng Wu}
\author[a]{Weihao Wu}
\author[a]{Jingkai Xia}
\author[n]{Mengjiao Xiao}
\author[d]{Xiang Xiao}
\author[g]{Pengwei Xie}
\author[a]{Binbin Yan}
\author[m]{Xiyu Yan}
\author[a]{Jijun Yang}
\author[a]{Yong Yang}
\author[h]{Chunxu Yu}
\author[e,f]{Jumin Yuan}
\author[a]{Ying Yuan}
\author[n]{Dan Zhang}
\author[a]{Minzhen Zhang}
\author[c]{Peng Zhang}
\author[a]{Tao Zhang}
\author[a]{Li Zhao}
\author[m]{Qibin Zheng}
\author[c]{Jifang Zhou}
\author[a]{Ning Zhou}
\author[k]{Xiaopeng Zhou}
\author[c]{Yong Zhou}

\affiliation[a]{School of Physics and Astronomy, Shanghai Jiao Tong University, MOE Key Laboratory for Particle Astrophysics and Cosmology, Shanghai Key Laboratory for Particle Physics and Cosmology, Shanghai 200240, China}
\affiliation[b]{Shanghai Jiao Tong University Sichuan Research Institute, Chengdu 610213, China}
\affiliation[c]{Yalong River Hydropower Development Company, Ltd., 288 Shuanglin Road, Chengdu 610051, China}
\affiliation[d]{School of Physics, Sun Yat-Sen University, Guangzhou 510275, China}
\affiliation[e]{Research Center for Particle Science and Technology, Institute of Frontier and Interdisciplinary Scienc, Shandong University, Qingdao 266237, Shandong, China}
\affiliation[f]{Key Laboratory of Particle Physics and Particle Irradiation of Ministry of Education, Shandong University, Qingdao 266237, Shandong, China}
\affiliation[g]{Tsung-Dao Lee Institute, Shanghai Jiao Tong University, Shanghai, 200240, China}
\affiliation[h]{School of Physics, Nankai University, Tianjin 300071, China}
\affiliation[i]{Key Laboratory of Nuclear Physics and Ion-beam Application (MOE), Institute of Modern Physics, Fudan University, Shanghai 200433, China}
\affiliation[j]{School of Physics, Peking University, Beijing 100871, China}
\affiliation[k]{School of Physics, Beihang University, Beijing 100191, China}
\affiliation[l]{International Research Center for Nuclei and Particles in the Cosmos \& Beijing Key Laboratory of Advanced Nuclear Materials and Physics, Beihang University, Beijing 100191, China}
\affiliation[m]{School of Medical Instrument and Food Engineering, University of Shanghai for Science and Technology, Shanghai 200093, China}
\affiliation[n]{Department of Physics, University of Maryland, College Park, Maryland 20742, USA}
\affiliation[o]{School of Mechanical Engineering, Shanghai Jiao Tong University, Shanghai 200240, China}
\affiliation[p]{State Key Laboratory of Particle Detection and Electronics, University of Science and Technology of China, Hefei 230026, China}
\affiliation[q]{Department of Modern Physics, University of Science and Technology of China, Hefei 230026, China}
\affiliation[r]{SJTU Paris Elite Institute of Technology, Shanghai Jiao Tong University, Shanghai, 200240, China}
\affiliation[s]{School of Physics and Microelectronics, Zhengzhou University, Zhengzhou, Henan 450001, China}

\emailAdd{mengyue@sjtu.edu.cn}
\emailAdd{siguang@pku.edu.cn}

\collaboration{PandaX-4T Collaboration}%
\abstract{PandaX-4T is a ton-scale dark matter direct detection experiment using a dual-phase TPC technique at the China Jinping Underground Laboratory. Various ultra-low background technologies have been developed and applied to material screening for PandaX-4T, including HPGe gamma spectroscopy, ICP-MS, NAA, radon emanation measurement system, krypton assay station, and alpha detection system. Low background materials were selected to assemble the detector. Surface treatment procedures were investigated to further suppress radioactive background. Combining measured results and Monte Carlo simulation, the total material background rates of PandaX-4T in the energy region of 1-25 keV$\rm{}_{ee}$ are estimated to be (9.9 $\pm$ 1.9) $\times \ 10^{-3}$ mDRU for electron recoil and (2.8 $\pm$ 0.6) $\times \ 10^{-4}$ mDRU for nuclear recoil. In addition, $^{nat}$Kr in the detector is estimated to be <8 ppt.}

\begin{document} 
\maketitle
\flushbottom

\section{Introduction}
A variety of astronomical and cosmological observations have implied that the universe contains more matter than ordinary matter, i.e. dark matter (DM) \cite{Bertone_2005}. Beyond the standard model, the Weakly Interacting Massive Particle (WIMP) is a competitive candidate for DM. Currently, dual-phase xenon detectors are leading in the search for WIMPs in a mass range from a few GeV/c$^{2}$ to several TeV/c$^{2}$ \cite{Juyal:2019gch,Giboni:2019fqo,Li:2021oos,Zhao:2020ezy}. The PandaX collaboration has constructed a series of dual-phase xenon time projection chambers (TPCs) to search for WIMPs at the China Jinping underground laboratory (CJPL)~\cite{Liu_2017}. PandaX-II provided that the lowest exclusion value on the spin-independent cross section is $2.2 \times 10^{-46}\ \rm{cm^{2}}$ (90\% C.L.) at a WIMP mass of 30 GeV/c$^{2}$ with 132 tonne$\cdot$day exposure in 2020 \cite{Wang_2020}. PandaX-4T located in the B2 hall of CJPL-II is now on commissioning and taking data \cite{meng2021dark,Zhang_2018}  with an active mass of 4 tonnes.

\begin{figure}[htbp]
    \centering
    \includegraphics[width=0.55\textwidth]{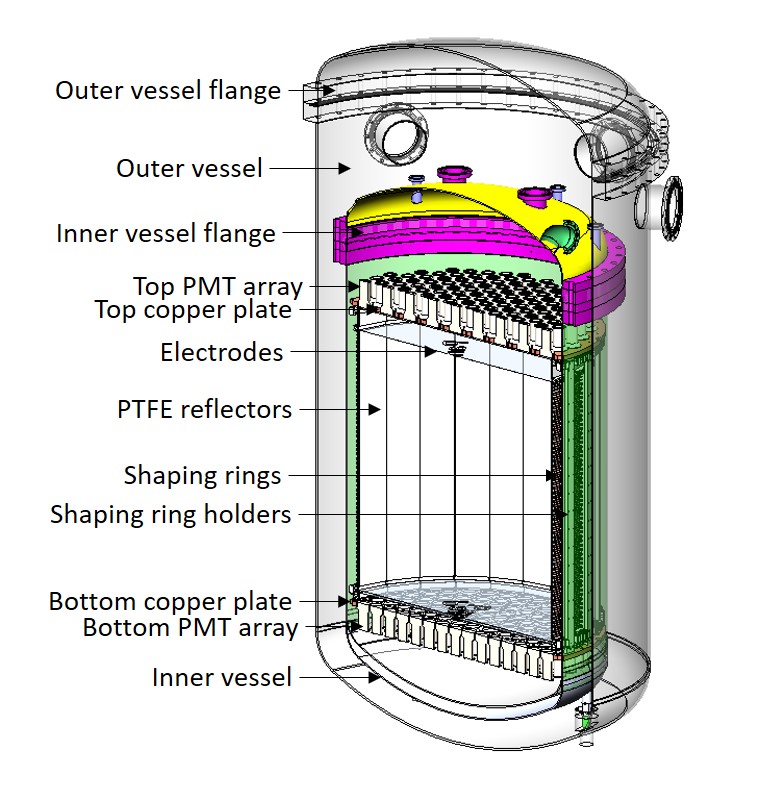}
    \caption{Layout of the PandaX-4T detector}
    \label{fig:layout}
\end{figure}

The PandaX-4T detector is placed in an ultra-pure water shielding with a height of 13 m and a radius of 5 m to reduce the external gammas and neutrons from the laboratory environment. The structure of PandaX-4T cryostat system and TPC are depicted in Figure~\ref{fig:layout}. The cryostat system is composed of an outer vessel (OV) and an inner vessel (IV) \cite{Zhao_2021,Zhang_2016}. Each vessel consists of a barrel, two domes, and a flange. The TPC, which is the core of the experiment, is placed inside the inner vessel and immersed in liquid xenon. Two arrays of photomultiplier tubes (PMTs) are set at the top and bottom of the TPC to collect photon signals. Copper plates are built to hold the PMT arrays. Between the copper plates there is the field cage. Shaping rings made of copper are placed around the TPC to maintain the drift electric field. Polytetrafluoroethylene (PTFE) reflectors are placed at the inside layer of the field cage to improve the efficiency of photon collection.

Radioactive background may mimic interaction signals between DM and ordinary matter. Thus, it is important to suppress, control, and better understand the background in DM experiments. PandaX-4T has developed and utilized various detection technologies to precisely measure the bulk radioactivity of materials, radon emanation, krypton concentration in the xenon target, and the radon daughters contamination on the surface. Thus, the total material background rate can be evaluated.

Section~\ref{sec:bkg_origins} introduces the main origins of the background contamination in PandaX-4T. Section~\ref{sec:counting} illustrates that the various techniques are established, the lowest radioactivity materials are selected to build the detector, and total background of PandaX-4T can be calculated with the results of the measurements. Section~\ref{sec:bkg_estimation} gives the final background estimation based on the screening results and simulation.
Detailed screening results of HPGe counting stations and summarized radioactivity input of background estimation are shown in the Appendix. 

\section{Radioactive Origins}
\label{sec:bkg_origins}
The background signals can be divided into two categories. Signals caused by interactions with atomic electrons are called electron recoil (ER) events, while signals resulted from interactions with the nuclei of the xenon target are nuclear recoil (NR) events \cite{Liu_2017,Undagoitia_2015}. 

Benefiting from the 2400-m rock overburden, the cosmic ray flux at CJPL can be suppressed to $3.53\times 10^{-10} \ \rm{cm^{-2}s^{-1}}$ \cite{JNE:2020bwn}. Thus, the muon-induced background is negligible. Therefore, the background of PandaX-4T originates from three main sources: radioactive isotopes in materials (bulk and surface), xenon target impurities, and neutrino-related background.

Long-lived radioactive isotopes such as $^{40}$K, $^{232}$Th, and $^{238}$U exist in all detector materials. Both their decays and subsequent decays from their daughter isotopes may result in background contributions. Also, $^{222}$Rn, a progeny of $^{238}$U decay, breaks the secular equilibrium of the $^{238}$U decay chain due to out-gassing radon which diffuses uniformly in the xenon target. Conversely, $^{220}$Rn from the $^{232}$Th chain can be neglected because of its short half-life (55 s). Radon daughters with longer half-life ($^{210}$Pb and $^{210}$Po) deposited on the detector surface may contribute to surface contamination. Industrial isotopes introduced during materials production like $^{60}$Co and $^{137}$Cs also lead to background contributions. For those isotopes, gamma and beta decays result in ER background. In addition, some isotopes in $^{232}$Th, $^{235}$U, and $^{238}$U decay chains release neutrons either through spontaneous fission or via ($\rm{\alpha}$, n) reactions, which leads to NR background events \cite{LZ_2020,XENON_2017,ngammaML}.  

Xenon target impurities also produce background events in the DM search energy range. For example, $^{85}$Kr is a beta emitter in commercial liquid xenon, which will result in ER background events.  

Neutrino-related background cannot be suppressed. $^{136}$Xe, a two-neutrino double-beta decay isotope comprising 8.9\% of natural xenon, makes ER background contributions. Solar neutrinos are irreducible so that related estimations need to be taken into consideration.

\section{Low Background Assay Techniques and Screening Results}
\label{sec:counting}
\begin{table*}[b]
\scriptsize
\centering
\caption{Sample materials used in PandaX-4T construction}
\label{table:parts}
\begin{tabular}{@{}llll@{}}
\toprule
Components & Quantities & Sample Name& Detector \\
  \hline
PMT R11410 & 368 pcs & PMT R11410 & JP-I  \\
Window of PMT R11410 & 368 pcs & Faceplate of PMT R11410 & JP-I \\
Stem of PMT R11410 & 368 pcs & Ceramic stem body of PMT R11410 & JP-I   \\
Base of PMT R11410  & 368 pcs& Base of PMT R11410 & JP-I, JP-II   \\
Spring of PMT R11410 & 1104 pcs& Spring of PMT R11410 & JP-I\\
PMT R8520 & 144 pcs & PMT R8520 & JP-II  \\
Base of PMT R8520& 144 pcs & Base of PMT R8520 & JP-II   \\
IV barrel and IV dome & 500 kg& P4TL & JP-I  \\
OV barrel& 960 kg & P4TJ & JP-I  \\
OV domes and electrodes& 400 kg & P4TI, P4TK & JP-I \\
Flanges and bolt screws &1.20 tonne& P4TG, P4TH & JP-I\\
Threaded insert & 8 pcs & Threaded insert & JP-I\\
Shaping rings and copper plates&200 kg & Copper \#1, Copper \#2 & ICP-MS  \\
PTFE holders and reflectors& 200 kg & PTFE & NAA  \\
\bottomrule
\end{tabular}
\end{table*}

To effectively determine the quantities of radioactivity from the material used in PandaX-4T construction and the impurity in the xenon target, PandaX-4T has established an ultra-low background platform, including two high purity germanium (HPGe) counting stations, inductively coupled plasma mass spectrometry (ICP-MS), neutron activation analysis (NAA), radon emanation measurement systems, krypton assay station, and alpha detection system. Batches of candidate samples used in PandaX-4T construction were measured and selected with these techniques. A summary of samples assayed and used in PandaX-4T is shown in Table \ref{table:parts}. During the raw material or parts production, processing, and transportation, the surface of materials used in detector construction may be contaminated due to oxidization, dust fall-out, and long-lived radon progenies ($^{210}$Pb and $^{210}$Po) deposition, resulting in increase of radioactivity on the surface of materials. Hence, PandaX-4T has investigated surface cleaning procedures and recipes in order to further reduce surface background contamination. In this section, radioassay techniques, their corresponding counting results, and surface cleaning procedures will be shown and discussed. 

\subsection{HPGe Counting Stations}
\begin{figure}[b]
\centering
\includegraphics[width=\linewidth]{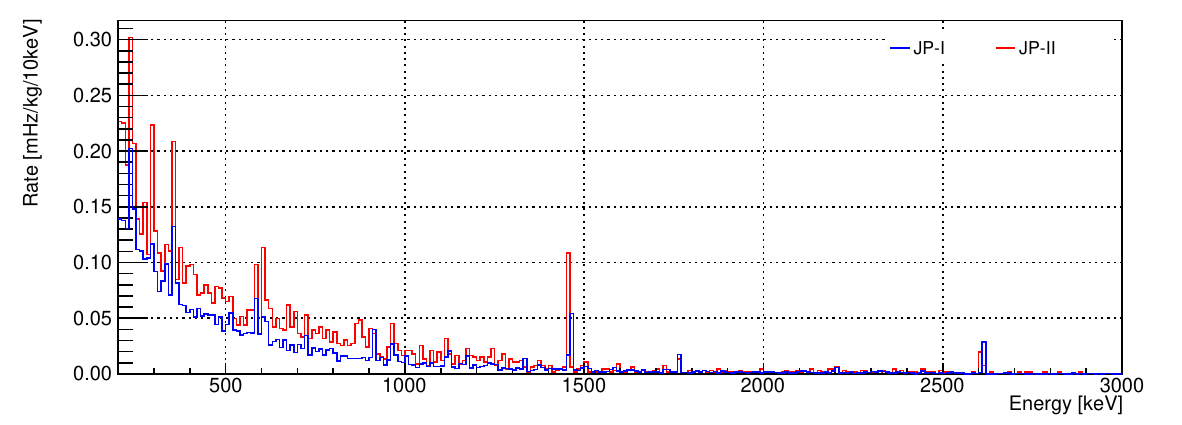}
\caption{\label{figure:HPGe_bkg}Background spectra of JP-I (blue line) and JP-II (red line) HPGe detectors. The counting rate is normalized by the mass of germanium crystal}
\end{figure}

\begin{figure}[t!]
\begin{minipage}[t]{0.49\textwidth}
\centering
\includegraphics[width=\linewidth]{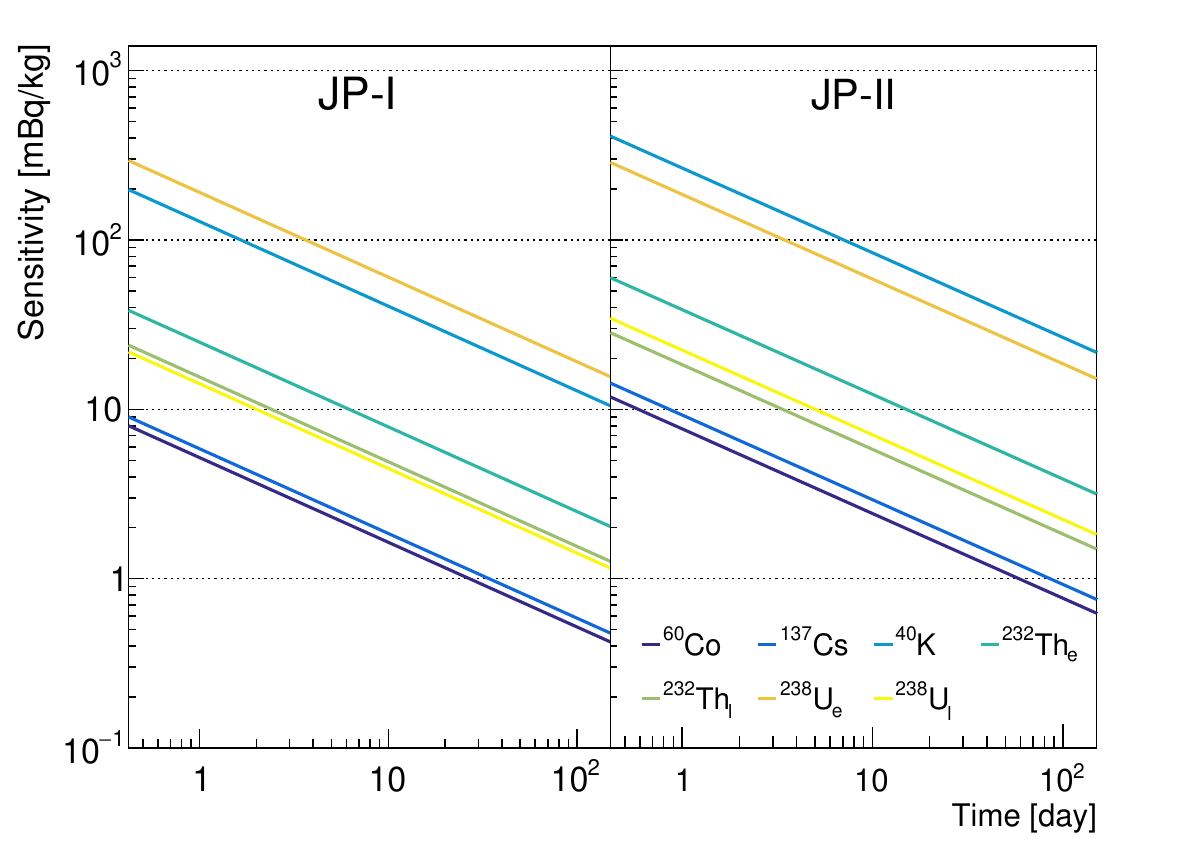}
\caption{\label{figure:sensitivity}Sensitivity of JP-I and JP-II HPGe detectors for a PTFE sample}
\end{minipage}\hspace{0.01\textwidth}
\begin{minipage}[t]{0.49\textwidth}
\centering
\includegraphics[width=\linewidth]{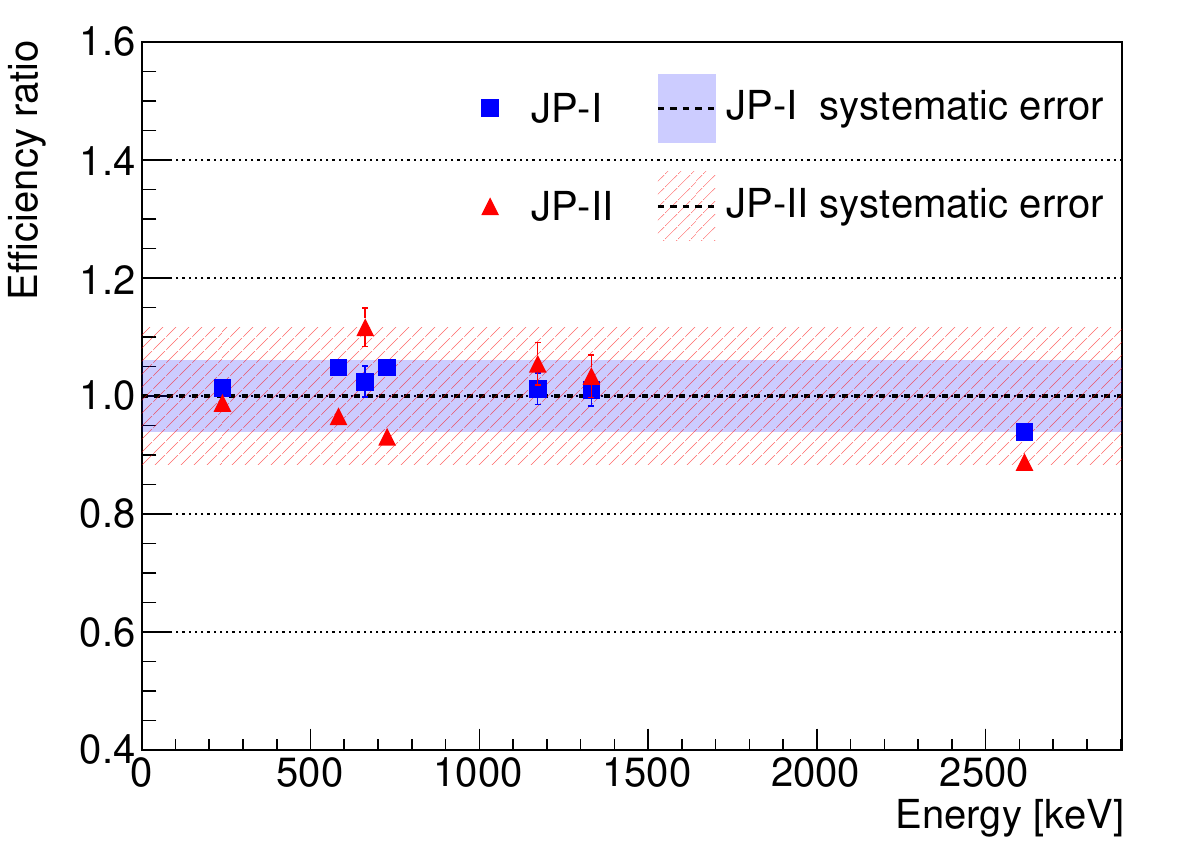}
\caption{\label{figure:cross-calibration} Efficiency ratios of measurement and simulation for JP-I and JP-II HPGe detectors}
\end{minipage} 
\end{figure}

At CJPL, two gamma counting stations (JP-I and JP-II in the following) are assembled with high-purity germanium crystals, a shielding structure with 10-cm-thick copper and 20-cm-thick lead to reject ambient gamma/neutron background, and a vacuum chamber in order to further reduce air radon \cite{Wang_2016,Yan:2020aif}. The background rate reaches 1.0 counts/min and 0.3 counts/min for JP-I and JP-II respectively at the energy range from 200 to 3000 keV. The background spectra are shown in Figure \ref{figure:HPGe_bkg}. Secular equilibrium breaks at $^{222}$Rn in the $^{238}$U decay chain. Hence, the radioactivity of the $^{238}$U early chain ($^{238}$U$_{e})$ and late chain ($^{238}$U$_{l}$) are measured separately. Similarly, the $^{232}$Th decay chain is split between $^{228}$Ac and $^{228}$Th. The characteristics of the two detectors are presented in Table \ref{table:HPGe_charac}. Figure \ref{figure:sensitivity} shows the sensitivity of JP-I and JP-II over duration for a typical measurement of PTFE sample with a mass of 173 g. The main reason for the sensitivity difference between the two detectors is the size of the germanium crystal.
A GEANT4-based \cite{GEANT4:2002zbu} simulation program is built to calculate detection efficiencies of each sample at given energies.

\begin{table}[htbp]
\scriptsize
\centering
\caption{Characteristics of the HPGe detectors}
\label{table:HPGe_charac}
\begin{tabular}{@{}lll@{}}
\toprule
Detector & JP-I & JP-II \\ \midrule
Manufacturer & ORTEC & Canberra \\
Type & P-type coaxial & Broad energy \\ 
Mass [kg] & 3.69 & 0.63 \\
Relative efficiency & 175\% & 34\% \\
Integral [200, 3000] keV [counts/min] & 1.0 & 0.3 \\
FWHM@ 662 keV [keV] & 2.4 & 2.7 \\
FWHM@ 1332 keV [keV] & 3.0 & 2.8 \\ \bottomrule
\end{tabular}%
\end{table}

Using three radioactive sources of $^{60}$Co, $^{137}$Cs, and $^{232}$Th with known activities and geometries, cross-calibration is performed to verify the consistency of the measurements of JP-I and JP-II. It can be obtained from Figure~\ref{figure:cross-calibration} that the efficiency ratios of JP-I and JP-II are around 1.0 in given energies. And 6\%/12\% systematic errors are assigned to the JP-I/JP-II detectors.

Screening results of JP-I and JP-II are listed in Table \ref{table:HPGe_results}, Appendix \ref{appendix:HPGe_results}. The uncertainties are given as $ \pm 1\sigma$ (statistical error) of detected activities or at 90\% confidence level (C.L.) for upper limits. All the samples are cleaned by certain cleaning methods before measurements. Generally, samples are wiped with alcohol and cleaned ultrasonically. Copper is treated by rigorous chemical procedures and PTFE is soaked with ultra-pure nitric acid. The samples measured with the HPGe detectors could be divided into three categories: metal, PMT components, and PTFE. Among them, the radioassay results of stainless steel (SS) and PMT components are used in the calculation of experiment background.

SS is widely used in detector construction, such as IV, OV, electrodes, etc. Hence, different batches of customized low-background SS from Taiyuan Iron and Steel Company (TISCO) are measured and selected. The batch number of SS finally used in PandaX-4T construction and parts applied correspondingly can be seen in Table \ref{table:parts}. PMT R11410, PMT R8520 from Hamamatsu and their components are also measured. It is worth mentioning that, capacitors, as the most radioactive electronic components of PMT bases, are carefully selected. From Table~\ref{table:HPGe_results}, capacitors from Yageo supplier are the least radioactive, followed by capacitors from Knowles supplier. Further considering trade-off of their electrical properties at low temperature and radioactivity, capacitors from Knowles were finally chosen.

\subsection{ICP-MS}
\label{subsec:icpms}
ICP-MS is a technology providing quick multi-elemental analysis with high sensitivity and wide atomic mass range. The intrinsic radioactivity (mainly $^{238}$U and $^{232}$Th) of sample materials can be indicated by measuring the concentration of long-lived radionuclides based on their mass-to-charge ratios. The ICP-MS assays for PandaX-4T were performed in a Class 10 cleanroom facility utilizing an Agilent 7900 spectrometer. In addition, an analytical balance, a sub-boiling distiller, a microwave digestion system, and an ultra-pure water system are equipped in the cleanroom. Different sample preparation methods are developed depending on the properties of samples, including dilution method and ion-exchange method with TEVA/UTEVA resins, etc. The details of copper and SS measurements can be found in \cite{YUAN:90301,FU:20502,Fu:2021}.

\begin{figure}[b!]
    \centering
    \includegraphics[width=0.6\textwidth]{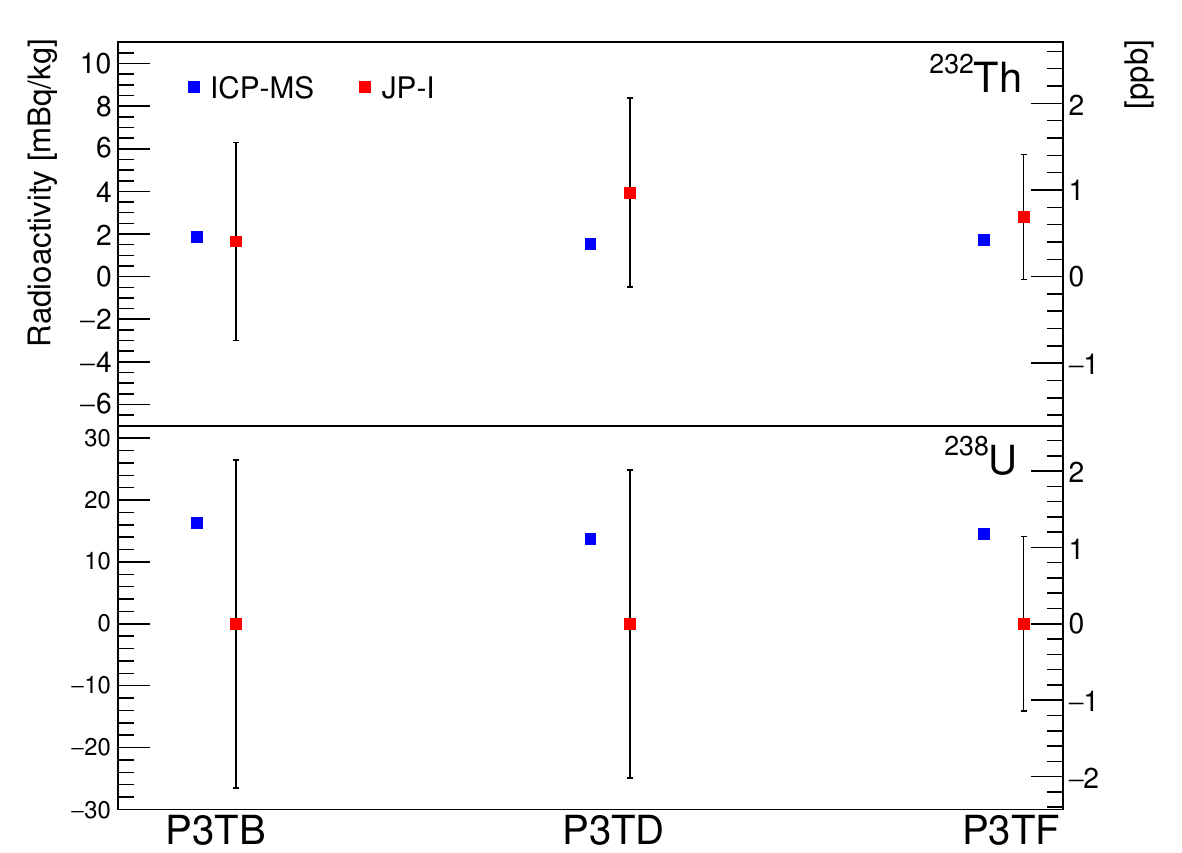}
    \caption{Cross validation for ICP-MS and JP-I HPGe for SS samples with batch number P3TB, P3TD, and P3TF}
    \label{fig:crosscheck_icpms}
\end{figure}

ICP-MS measures the $^{238}$U and $^{232}$Th content directly, which should be consistent with the radioactivity of $^{238}$U$_{e}$ and $^{232}$Th$_{e}$ measured by HPGe detectors. Cross validation with ICP-MS and HPGe using three batches of SS (batch number: P3TB, P3TD, and P3TF) from TISCO is performed. In these ICP-MS measurements, the SS samples are prepared using the dilution method \cite{Fu:2021}. By comparing the measurement results in Table \ref{table:ICPMS_results} and Table \ref{table:HPGe_results}, it can be concluded that they are consistent with each other within uncertainties (shown in Figure \ref{fig:crosscheck_icpms}).


\begin{table}[t]
\scriptsize
\centering
\caption{Investigation of pickling recipes to remove $^{238}$U and $^{232}$Th on copper surface. The results show the relative $^{238}$U and $^{232}$Th change with dissolve a skin layer of copper.}
\label{table:Cu_cleaning}
\begin{tabular}{clcc}
\toprule
No. & Pickling recipes & $^{232}$Th [mBq/kg] & $^{238}$U [mBq/kg] \\ \hline
1 &None & 25.62 $\pm$ 0.66 & 64.78 $\pm$ 3.14 \\
2 &Touch   repeatedly by hand & 83.08 $\pm$ 1.37 & 78.10 $\pm$ 3.21 \\
3 & 1\% H$_2$SO$_4$ + 3\% H$_2$O$_2$ & 5.54 $\pm$ 0.07 & 21.94 $\pm$ 0.49 \\
4 & 15\% HNO$_3$ + 2\% H$_2$O$_2$ & 22.22 $\pm$ 0.14 & 44.63 $\pm$ 1.14 \\
5 & 5\% C$_6$H$_8$O$_7$ + 15.7\% H$_2$O$_2$ & 12.17 $\pm$ 0.30 & 48.88 $\pm$ 1.25\\ \bottomrule
\end{tabular}%
\end{table}


ICP-MS can not only measure bulk radioactivity, but also surface radioactivity by dissolving skin layer of samples during preparation procedures. Since detector components like shaping rings and copper plates inside TPC are very close to the central xenon target, cleaning procedures which aims at reducing surface $^{238}$U and $^{232}$Th content for copper were carried out. The procedure consists of four steps: degrease, pickling, passivation and drying. The first step is to immerse the sample in acetone for ultrasonic cleaning (15 mins), and then in alcohol for another 15 minutes. Second, different acid solutions H$_2$SO$_4$, HNO$_3$, and citric acid (C$_6$H$_8$O$_7$) were tested. Third, samples were immersed in 1\% C$_6$H$_8$O$_7$ for passivation. Last, blow the samples with nitrogen gas and bake at 65 $^{\circ}$C for 30 mins. The results of different pickling solution are shown in Table \ref{table:Cu_cleaning}. The first two measurements serve as control groups. It can be concluded that all three cleaning procedures have effect on surface radioactivity reduction, while the pickling recipe of 1\% H$_2$SO$_4$ + 3\% H$_2$O$_2$ performs best among them.

\begin{table}[t]\scriptsize
\centering
\caption{ICP-MS measurement results (upper limit with 90\% C.L. is given)}
\label{table:ICPMS_results}
\begin{tabular}{@{}llllll@{}}
\toprule
Name                     & Supplier & Material        & unit   & $^{232}$Th                & $^{238}$U             \\
\hline
\textbf{Metal} \\
P3TB                     & TISCO    & Stainless steel & mBq/kg & 1.86 $\pm$ 0.24            & 13.34 $\pm$ 0.48       \\
P3TD                     & TISCO    & Stainless steel & mBq/kg & 1.52 $\pm$ 0.25            & 13.66 $\pm$ 0.72       \\
P3TF                     & TISCO    & Stainless steel & mBq/kg & 1.73 $\pm$ 0.19            & 14.54 $\pm$ 0.41       \\
Copper \#1               & LUOYANG  & Copper          & ppt\footnotemark[1]    & 1.27 $\pm$ 0.34            & 6.20 $\pm$ 0.32        \\
Copper \#2               & LUOYANG  & Copper          & ppt    & \textless{}0.55      & 2.85 $\pm$ 0.21        \\ \\
\hline
\textbf{Water}\\
Underground water \#1    & -        & water           & ppt    & 0.28 $\pm$ 0.04            & 142.06 $\pm$ 0.60      \\
Underground water \#2    & -        & water           & ppt    & 0.39 $\pm$ 0.06            & 81.60 $\pm$ 3.20        \\
Ultra-pure water               & -        & water           & ppt    & 0.06 $\pm$ 0.02            & (3.00 $\pm$ 0.10)$\times10^{-2}$ \\
Storage water (1 day)    & -        & water           & ppt    & 0.14 $\pm$ 0.03            & 0.35 $\pm$ 0.01        \\
Storage water (7 days)   & -        & water           & ppt    & 0.14 $\pm$ 0.02            & 0.32 $\pm$ 0.05        \\
Storage water (15 days)  & -        & water           & ppt    & 0.14 $\pm$ 0.01            & 0.27 $\pm$ 0.01        \\

\bottomrule
\end{tabular}%
\end{table}

\footnotetext[1]{1 ppt is equal to 0.004 mBq/kg for $^{232}$Th and 0.012 mBq/kg for $^{238}$U. }

Screening results of ICP-MS are presented in Table \ref{table:ICPMS_results}. As mentioned before, copper is another important metal used in PandaX-4T detector. Low background copper from LUOYANG is chosen. Copper samples are processed with the cleaning procedure mentioned above before measurements. ICP-MS gives precise measurements and the results are applied to the background calculation of the experiment. The JP-I HPGe also provides a measurement for copper, while it can only give an upper limit due to copper's low radioactivity (Table \ref{table:HPGe_results}). Various water samples which are used for water shielding were measured to examine the effect of the purification system and long-term water soaking contamination. It can be obtained that the radioactivity of purified water is 4 orders of magnitude less than that of underground water for $^{238}$U and 10 times less for $^{232}$Th. Meanwhile, the radioactivity of ultra-pure water in the shielding tank did not change much over time, which indicates no obvious surface contamination exists from SS water shielding tank within 15-day soaking time. With Monte Carlo simulation, the background due to the intrinsic $^{238}$U and $^{232}$Th in the water can be neglected ($\sim10^{-7}$ mDRU). Thus, the purified water is able to serve as water shielding to reduce external gammas and neutrons.

\subsection{Radon Emanation systems}
\begin{table}[b]
\scriptsize
\centering
\caption{Performance of radon emanation measurement systems.}
\label{table:Rn_systems}
\begin{tabular}{c c c c c c}
\toprule
No. & Chamber &  Cold trap & Volume [L] & Blank [mBq] & Efficiency [\%] \\
\hline
1 & Cylinder & no & 7.4& 1.07 $\pm$ 0.01 & 24.6 $\pm$ 0.2\\
2 & T-type & no & 0.7& 0.08 $\pm$ 0.01 & 12.7 $\pm$ 1.0\\
3 & T-type & yes & 0.7& <0.10 & 7.9 $\pm$ 0.1\\
\bottomrule
\end{tabular}
\end{table}

\begin{figure}[b!]
\centering
\begin{subfigure}[t]{0.25\textwidth}
\centering
\includegraphics[width=\linewidth]{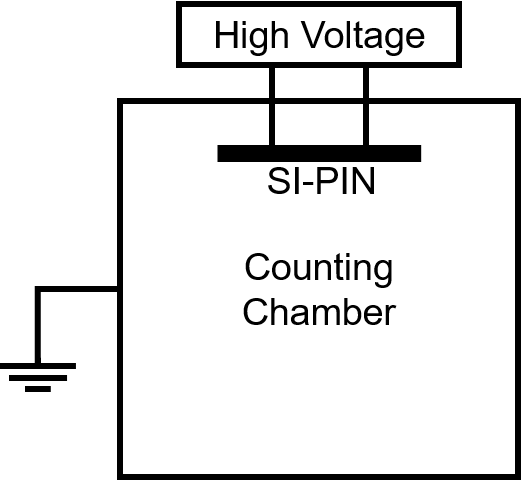}
\caption{\label{figure:distribution}System 1}
\end{subfigure}\hfill
\begin{subfigure}[t]{0.65\textwidth}
\centering
\includegraphics[width=\linewidth]{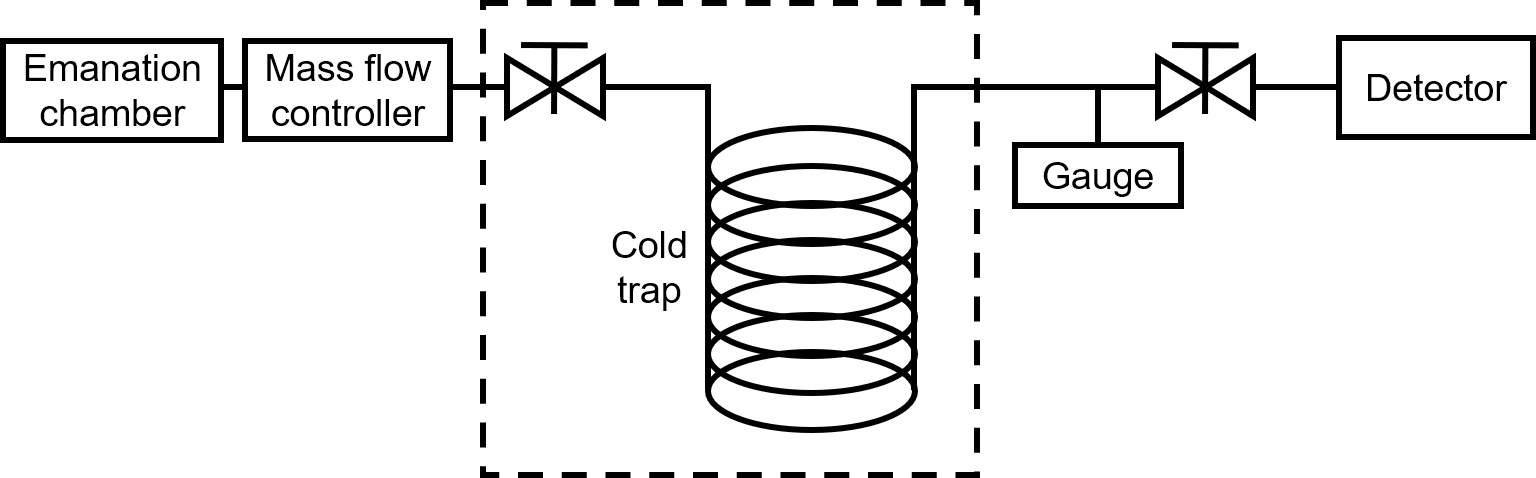}
\caption{\label{figure:s_FVMass} System 2 $\And$ 3 (The difference between system 2 and 3 is the cold trap.)}
\end{subfigure} 
\caption{Schematics of radon emanation systems.}
\label{figure:rn_layout}
\end{figure}

The $^{222}$Rn emanation rate can be measured by the decay rates of the radon daughter isotopes, such as $^{218}$Po ($\rm{t}_{1/2}=3.1$ mins) and $^{214}$Po ($\rm{t}_{1/2}=164\ \rm{\mu s}$) \cite{Li_2017,Miller:2017tpl}. The PandaX-4T collaboration has designed three radon emanation measurement systems. Table \ref{table:Rn_systems} lists the details and Figure \ref{figure:rn_layout} shows the schematics of the three systems. These systems use an electrostatic method to collect the positively charged ions of radon daughter nuclei under electric field and use SI-PIN diodes to measure energy deposition. The efficiency of each system is calibrated using a $^{226}$Ra radioactive source with known activity, which includes both the detection and transfer efficiencies. A cold trap at liquid nitrogen temperature can be added to the system to achieve $^{222}$Rn enrichment. 360 L gas emanated from samples could be collected with a cold trap at a flow rate of 1 slpm for six hours to achieve $^{222}$Rn enrichment by a factor of 500. Depending on the properties of the sample to be tested, different systems are chosen to make appropriate measurements.

The radon emanation of the following subsystems in PandaX-4T and samples is measured (Table \ref{table:Rn_result}). The radon emanation rate of the PMT R11410 is obtained by scaling surface area from the measurement of the PMT R12699 with the assumption that the materials are similar for both.
During operation, radon emanated from the distillation tower (DT), which is used to remove $^{85}$Kr, will enter the xenon target since it is connected to the TPC. For the cryostat system, radon gas released by the inner vessel will directly get into the TPC. The getter, which is designed to remove electronegative gas (e.g., O$_2$, N$_2$) in the xenon, is measured under room temperature. The KNF pump, which is used to circulate the whole system, is also measured under different operation conditions to monitor if extra $^{222}$Rn will be induced. Zirconium (Zr) beans is a candidate material to remove tritium in xenon target. The $^{222}$Rn level in PandaX-4T detector can be better controlled with the screening results.

\begin{table}[tbp]
\scriptsize 
\centering
\caption{Radon emanation measurement results (upper limit with 90\% C.L. is given). The serial number of detector corresponds to Table \ref{table:Rn_systems}. All the measurement are performed at room temperature.}
\label{table:Rn_result}
\begin{tabular}{@{}lllll@{}}
\toprule
Name                                         & Supplier & Detector & Unit   & Rate             \\ \hline
PMT R11410& Hamamatsu & 1 & mBq/pc & \textless{0.03}\\
Distillation tower                           & -        & 3        & mBq    & 19.4 $\pm$ 5.3 \\
Inner vessel                                 & -        & 3        & mBq    & \textless{}17.9    \\
Getter                                      & SAES     & 2        & mBq    & \textless{}0.4  \\
Getter                                     & SimPure  & 2        & mBq    & \textless{}0.4  \\
Diaphragm pump running (exposed to air) & KNF      & 2        & mBq    & 3.1 $\pm$ 0.5        \\
Diaphragm pump not running (wrapped)    & KNF      & 2        & mBq    & 1.7 $\pm$ 0.3        \\
Diaphragm                                    & KNF      & 1        & mBq/pc & \textless{}0.02  \\ 
Zr beans                                 & NANJING YOUTIAN        & 1        & mBq/m$^{2}$ & \textless{}13.06 \\ \bottomrule
\end{tabular}%
\end{table}

\subsection{Krypton Assay Station}
\begin{figure}[b]
    \centering
    \includegraphics[width=0.8\textwidth]{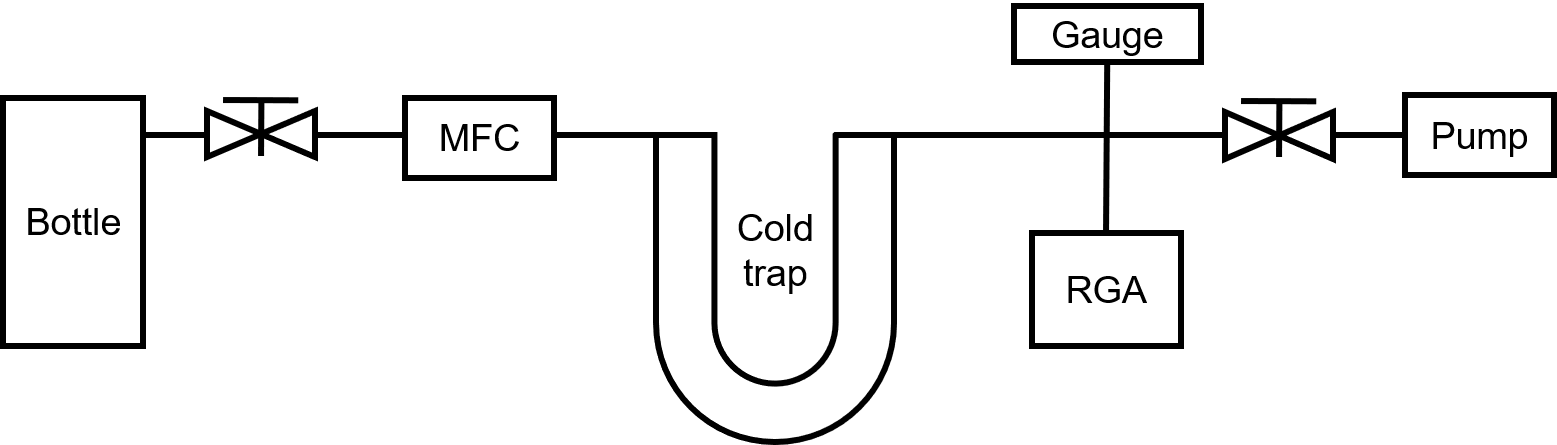}
    \caption{Layout of krypton assay station}
    \label{fig:layout_Kr}
\end{figure}
The krypton assay station is established at CJPL to measure the krypton concentration in xenon target. Figure \ref{fig:layout_Kr} is the schematic drawing of the krypton assay station, which shows that the main components of the system are a residual gas analyzer (RGA), a cold trap, mass flow controllers (MFCs), a cold gauge, bottles, and pumps. The typical sensitivity of a RGA is $\sim$1 ppm. To measure sample gas whose Kr-Xe concentration ratio is lower than 1/10$^{6}$, a cold trap is added to the system. Making use of the difference of their vapor pressures, the cold trap is able to separate krypton and xenon. This is equivalent to enriching the $^{nat}$Kr ratio by a factor of 10$^{5}$. The sensitivity of the $^{nat}$Kr assay station can then be improved to $\sim$10 ppt \cite{Cui:2020bwf}. The station can not only measure krypton concentration in the xenon target, but also monitor and evaluate the xenon purification with DT and the out-gassing from material.

Several groups of $^{nat}$Kr assay results under different conditions are shown in Table \ref{table:Kr_result}. After purification, the krypton concentration can be theoretically reduced from $5\times10^{-7}$ mol/mol to $\sim10^{-14}$ mol/mol \cite{Cui:2020bwf}. One of the original xenon measurement results is shown in the table \ref{table:Kr_result}. The third and fourth groups show that no evidence of krypton impurities brought in both liquid and gas xenon during operation. Based on the measurements, the concentration of $^{nat}$Kr in PandaX-4T is estimated to be <8 ppt. The natural abundance of $^{85}$Kr in $^{nat}$Kr is $2\times10^{-11}$. Thus, the concentration of $^{85}$Kr can be calculated.

\begin{table}[h]
\scriptsize
\centering
\caption{$^{nat}$Kr concentration measurement results (upper limit with 90\% C.L. is given) }
\label{table:Kr_result}
\begin{tabular}{@{}clc@{}}
\toprule
No.&Name  & Concentration [ppt]\\
\hline
1&Original xenon  & 1953 $\pm$ 130 \\
2&Product xenon  & \textless{8} \\
3&Liquid xenon sampling during operation  & \textless{12} \\
4&Gas xenon sampling during operation  & \textless{26} \\
\bottomrule
\end{tabular}%
\end{table}

\subsection{Alpha Detection System}
The alpha detection system aims at measuring the surface radioactivity of the sample material. A commercial alpha detection system fabricated by ORTEC is employed. It uses an ion-implanted-silicon charged-particle radiation detector of the ULTRA-AS series from ORTEC to catch alpha radiations from the surface of sample materials, mainly $^{210}$Po. The vacuum of the 966 cm$^{3}$ chamber can reach 10 mTorr and the blank is about 0.2 mHz. A Monte Carlo simulation based on the GEANT4 toolkit \cite{GEANT4:2002zbu} was used to calculate the detection efficiency for each sample. 

In order to remove radon daughters on copper surface, efficient surface cleaning methods have been investigated with the alpha detection system \cite{FU:20502}. Similar to the cleaning procedures to remove surface $^{238}$U and $^{232}$Th described in Section \ref{subsec:icpms}, the following four steps are included: degrease, pickling, passivation, and drying. For the first step, immerse the sample in 1\% Alconox detergent and clean ultrasonically for 20 mins. The other steps are the same as previous description. Each group of pickling solution was measured three times and the average removal efficiency of $^{210}$Po is demonstrated in Table \ref{table:alpha}. It can be seen that citric acid is the most effective.

\begin{table}[htbp]
\scriptsize
\centering
\caption{Cleaning procedure investigation to remove radon daughters on copper surface \cite{FU:20502}}
\label{table:alpha}
\begin{tabular}{@{}clc@{}}
\toprule
No. & Pickling solution & Average Removal Efficiency {[}\%{]} \\ \midrule
1 & 1\% H$_2$SO$_4$ + 3\% H$_2$O$_2$ & 79.4 $\pm$ 2.7 \\
2 & 15\% HNO$_3$ + 2\% H$_2$O$_2$ & 15.2 $\pm$ 2.3 \\
3 & 5\% C$_6$H$_8$O$_7$ + 8\% H$_2$O$_2$ & 99.9 $\pm$ 2.3 \\ \bottomrule
\end{tabular}%
\end{table}

\subsection{NAA}
Neutron activation analysis refers to a technique of analyzing elemental composition of a sample material using neutron activation. In the sample material, nuclei activated by neutrons emit characteristics $\gamma$ radiation and decay into corresponding daughter nuclei. Take $^{238}$U as an example, the reaction chain is shown as follows:
\begin{equation}
    ^1\rm{n} + ^{238}\rm{U} \xrightarrow[]{} \ ^{239}\rm{U} \xrightarrow[23.5\ \rm{mins}]{\beta^-} \ ^{239}\rm{Np} \xrightarrow[2.36\ d]{\beta^-} \ ^{239}\rm{Pu}
\end{equation}
$^{238}$U activated by neutrons decays to $^{239}$Pu with sufficiently long half lives ($2.4\times10^4$ y). The decay time of activated atoms can be boosted by a factor of $10^{11}$ (= ratio of half-lives of $^{238}$U and $^{239}$Np). The gamma radiation occurring during the reaction could be thus measured by a HPGe detector. NAA can achieve substantially greater sensitivity than direct $\gamma$ ray counting, typically at ppt and sub-ppt levels \cite{junocollaboration2021radioactivity}. 

For PandaX-4T, NAA was used for screening PTFE, using the TRIGA Mark II research reactor of the University of Pavia (Italy) as neutron source and HPGe detectors at the Radioactivity Laboratory of Milano-Bicocca University. The raw materials of PTFE are provided by SANXIN and the processing is done by DONGYUE. The measurement result of PTFE by NAA is shown in Table \ref{table:NAA}. Direct measurements of four batches of PTFE raw materials and two batches of processed PMT R8520 holders by JP-I are carried out as well (Table \ref{table:HPGe_results}). In addition, PTFE parts were soaked in 35\% HNO$_3$ for 1 week before the assembly of the detector to reduce surface contamination. 

\begin{table}[h]
\scriptsize
    \centering
    \caption{NAA measurement of PTFE for PandaX-4T (unit: mBq/kg)}
    \label{table:NAA}
    \begin{tabular}{lp{2.5cm}<{\centering}p{2.5cm}<{\centering}p{2.5cm}<{\centering}}
    \toprule
    Name & $^{40}$K & $^{232}$Th & $^{238}$U\\
    \hline
    PTFE & $(4.65 \pm 0.19)\times 10^{-5}$ & 0.04 $\pm$ 0.01 & \textless{0.01} \\
    \bottomrule 
    \end{tabular}
\end{table}

\section{Background Estimation}
\label{sec:bkg_estimation}
The background estimation of PandaX-4T experiment greatly depends on Geant4 simulation. A program called BambooMC \cite{Chen:2021asx} is specially designed to simulate the physical processes in the PandaX-4T detector.

Four cuts are developed in simulation to select DM candidates: single-scatter cut, veto cut, energy region of interest (ROI) cut, and fiducial volume (FV) cut.

\begin{itemize}
    \item Single-scatter cut\\
    According to the WIMPs model, the possibility of a DM particle scattering more than once with the xenon target is extremely small and can be neglected. The single scatter cut rejects events with multiple energy depositions in the simulation.
    \item Veto cut\\
    In BambooMC, a skin layer is designed corresponding to the veto compartment containing PMT R8520 in the detector. In this skin layer, the events that deposit energy more than 705 keV$\rm{}_{ee}$ are rejected.
    \item ROI cut\\
    The ROI of dark matter is set to be [1, 25] keV$\rm{}_{ee}$.
    \item FV cut\\
    Due to the self-shielding effect of liquid xenon, the vast majority of background events from materials deposit energy near the edge of the xenon target. Hence, the fiducial volume is necessary to be chosen in order to reject background events and increase the sensitivity (signal-noise ratio). In this result, the same FV volume as in \cite{meng2021dark} is chosen.
\end{itemize}

\begin{table}[h]
\scriptsize
\centering
\caption{Material background contribution in PandaX-4T detector (unit: mDRU)}
\label{table:mDRU}
\begin{tabular}{@{}lcc@{}}
\toprule
Unit: mDRU & ER & NR \\
\hline
PMT & (5.3 $\pm$ 1.2) $\times \ 10^{-3}$ & (8.9 $\pm$ 1.5) $\times \ 10^{-5}$\\
PTFE & (2.1 $\pm$ 0.3) $\times \ 10^{-5}$ & (8.2 $\pm$ 1.3) $\times \ 10^{-6}$\\
Copper & (1.7 $\pm$ 0.2) $\times \ 10^{-6}$& (2.5 $\pm$ 0.2) $\times \ 10^{-8}$\\
Inner vessel & (1.9 $\pm$ 0.8) $\times \ 10^{-3}$&(4.7 $\pm$ 3.7) $\times \ 10^{-5}$ \\
Outer vessel & (2.7 $\pm$ 1.3) $\times \ 10^{-3}$ & (1.4 $\pm$ 0.5) $\times \ 10^{-4}$\\
\hline
\textbf{Total Material} & \textbf{(9.9 $\pm$ 1.9) $\times \ 10^{-3}$} & \textbf{(2.8 $\pm$ 0.6) $\times \ 10^{-4}$}\\
\bottomrule
\end{tabular}
\end{table}

Combining the radioactive assay results (summarized in Table \ref{table:input}) and Monte Carlo simulation, the total material background of PandaX-4T can be estimated. The non-uniformity of the radioactive isotopes distribution in the material is not considered in this calculation. The background rate is calculated in units of mDRU (= $10^{-3}$ events kg$^{-1}$day$^{-1}$keV$^{-1}$). The background contributions from different components are summarized in Table \ref{table:mDRU}.
PMT, inner vessel, and outer vessel are the three major background sources.

\section{Summary}
The PandaX-4T facility is ready since August, 2019. Various screening technologies have been established and are able to meet the requirements of different radioactivity measurements. Strict material assays were performed to minimize the material background. Based on the counting results, the total material background of PandaX-4T is calculated to be (9.9 $\pm$ 1.9) $\times \ 10^{-3}$ mDRU for ER and (2.8 $\pm$ 0.6) $\times \ 10^{-4}$ mDRU for NR. In addition, $^{nat}$Kr in the detector is estimated to be \textless{} 8 ppt. Meanwhile, surface cleaning procedures were investigated to further reduce the material surface background, and to make sure that no extra contamination will be introduced during the transport and assembly process.
Relative to PandaX-II \cite{Wang_2020}, the rigorous selection of materials and cleaning procedures lead to a reduction of $\sim$95\% (87\%) in material background for ER (NR).
The $^{222}$Rn level and the $^{nat}$Kr concentration obtained from the commissioning run is 5.9 $\pm$ 0.1 $\mu$Bq/kg and 0.33 $\pm$ 0.21 ppt \cite{meng2021dark}, which decreased $\sim$40\% and 96\% compared to PandaX-II. 

The sensitivity of WIMP-nucleon interaction is largely dependent on the background of the detector. 
The sensitivity is proportional to exposure time, fiducial volume mass, and inversely proportional to square root of background rate.
Suppressing the material background helps extend fiducial volume and reduce total background rate of the detector, leading to an improvement of sensitivity. With the background level estimated in this paper and the PandaX-4T commissioning run data set, the lowest excluded value of the dark matter-nucleon spin-independent interactions reaches $3.8\times 10^{-47}$ cm$^2$ at 40 GeV/c$^2$ with an exposure of 0.63 tonne$\cdot$year [7]. Besides WIMP searching, the PandaX-4T experiment has many other physical goals, such as measurement of neutrinoless double beta decay (NLDBD) half-life. The majority of the background events in the ROI ([2.4, 2.7] MeV) of NLDBD is from detector components and radon emanation. Lower the material background and enlarge the fiducial volume at the same time can boost the sensitivity to measure the half-life.
\\
\\
\acknowledgments
We thank Monica Sisti in Milano-Bicocca University to help on NAA measurement. This project is supported by a grant from the Ministry of Science and Technology of China (No. 2016YFA0400301), grants from National Science Foundation of China (Nos. 12090060, 12005131, 11905128, 11925502, 11775141), Office of Science and Technology, Shanghai Municipal Government (No. 18JC1410200), Shanghai Pujiang Program (No. 19PJ1405800), Sichuan Science and Technology Program (No. 2020YFSY0057) and a grant from the Ministry of Science and Technology of China (2019YFE0114300). We also thank supports from Double First Class Plan of the Shanghai Jiao Tong University, sponsorship from Organization Department of Sichuan Providence, Department of Human Resources and Social Security of Sichuan Providence, Talent Recruiting Office of Tianfu New Area in Sichuan Providence, the Chinese Academy of Sciences Center for Excellence in Particle Physics (CCEPP), Hongwen Foundation in Hong Kong, and Tencent Foundation in China. Finally, we thank the CJPL administration and the Yalong River Hydropower Development Company Ltd. for indispensable logistical support and other help.

\bibliography{ref.bib}

\newpage
\appendix
\section*{Appendix}

\section*{Supplement to radioactive measured results}
\label{appendix:HPGe_results}
In the appendix, the radioactive supplement results are listed. The sample screening results with two HPGe detectors are shown in Table~\ref{table:HPGe_results}. Batch material screening was performed and low radioactive materials were selected for the PandaX-4T detector construction. Table~\ref{table:input} summarizes the radioactivity inputs (including statistical errors and systematic errors) of background estimation.

\clearpage
\thispagestyle{empty}

\begin{sidewaystable*}[tp]
\centering
\caption{HPGe counting station measurement results (upper limit with 90\% C.L. is given)}
\label{table:HPGe_results}
\begin{adjustbox}{width=\textwidth,center}
\begin{tabular}{l l l l l l c c c c c c c c}
\toprule
Name        & Supplier       & Detector & Material        & Mass (kg) & Unit   & $^{60}$Co             & $^{137}$Cs            & $^{40}$K               & $^{232}$Th$_{e}$           & $^{232}$Th$_{l}$           & $^{235}$U             & $^{238}$U$_{e}$             & $^{238}$U$_{l}$            \\ \hline
\textbf{Metal} \\
P3TB & TISCO & JP-I & Stainless steel & 1.24 & mBq/kg & \textless{}3.59 & \textless{}3.24 & \textless{}29.10 & \textless{}9.28 & \textless{}3.52 & \textless{}17.00 & \textless{}43.55 & \textless{}4.65 \\
P3TD & TISCO & JP-I & Stainless steel & 0.77 & mBq/kg & \textless{}3.02 & \textless{}2.89 & \textless{}37.62 & \textless{}11.23 & \textless{}8.62 & \textless{}4.43 & \textless{}42.65 & \textless{}4.50 \\
P3TF & TISCO & JP-I & Stainless steel & 0.83 & mBq/kg & \textless{}1.99 & \textless{}1.97 & \textless{}44.36 & \textless{}7.63 & 6.62 $\pm$ 2.01 & \textless{}7.38 & \textless{}23.16 & \textless{}3.93 \\
P4TD        & TISCO          & JP-I    & Stainless steel & 2.31      & mBq/kg & \textless{}2.27 & \textless{}2.26 & \textless{}60.60 & \textless{}6.47  & \textless{}3.07  & \textless{}7.81  & \textless{}44.04  & \textless{}4.02  \\
P4TG        & TISCO          & JP-I    & Stainless steel & 0.84      & mBq/kg & \textless{}1.77 & \textless{}1.69 & \textless{}33.12 & \textless{}5.69  & \textless{}5.16  & \textless{}9.92  & \textless{}23.64  & \textless{}1.92  \\
P4TH        & TISCO          & JP-I    & Stainless steel & 0.82      & mBq/kg & \textless{}2.32 & \textless{}2.29 & \textless{}41.37 & \textless{}8.85  & 8.11 $\pm$ 2.34       & \textless{}5.69 & \textless{}27.97  & \textless{}6.78 \\
P4TI        & TISCO          & JP-I    & Stainless steel & 0.87      & mBq/kg & \textless{}3.00 & \textless{}2.47 & \textless{}34.76 & \textless{}12.20 & \textless{}10.07 & \textless{}7.30  & \textless{}81.37  & \textless{}4.51  \\
P4TJ        & TISCO          & JP-I    & Stainless steel & 0.64      & mBq/kg & \textless{}1.61 & \textless{}1.87 & \textless{}28.56 & \textless{}7.76  & \textless{}5.42  & \textless{}6.63  & \textless{}32.56  & \textless{}3.63  \\
P4TK        & TISCO          & JP-I    & Stainless steel & 1.53      & mBq/kg & \textless{}2.37 & \textless{}1.31 & \textless{}17.57 & \textless{}4.60  & \textless{}2.05  & \textless{}1.94  & \textless{}29.07  & \textless{}2.16  \\
P4TL        & TISCO          & JP-I    & Stainless steel & 1.45      & mBq/kg & \textless{}3.13 & \textless{}2.28 & \textless{}34.42 & \textless{}7.64  & \textless{}3.66  & \textless{}4.74  & \textless{}95.20  & \textless{}4.56  \\
P4TP        & TISCO          & JP-I    & Stainless steel & 0.70      & mBq/kg & \textless{}2.78 & \textless{}2.73 & \textless{}39.04 & \textless{}10.34 & \textless{}2.92  & \textless{}4.32  & \textless{}44.55  & \textless{}6.60  \\
P4TQ        & TISCO          & JP-I    & Stainless steel & 1.05      & mBq/kg & \textless{}3.39 & \textless{}3.39 & \textless{}33.75 & \textless{}9.16  & \textless{}10.28 & \textless{}19.65 & \textless{}102.51 & \textless{}10.86 \\
P4TR        & TISCO          & JP-I    & Stainless steel & 0.92      & mBq/kg & \textless{}2.35 & \textless{}2.48 & \textless{}51.76 & \textless{}7.52  & \textless{}8.96  & \textless{}4.18  & \textless{}88.84  & \textless{}6.70  \\
P4TS        & TISCO          & JP-I    & Stainless steel & 0.66      & mBq/kg & \textless{}5.73 & \textless{}5.18 & \textless{}88.22 & \textless{}18.72 & \textless{}16.24 & \textless{}8.61  & \textless{}105.37 & \textless{}17.37 \\
P4TT        & TISCO          & JP-I    & Stainless steel & 1.00      & mBq/kg & \textless{}2.56 & \textless{}2.49 & \textless{}44.01 & \textless{}7.98  & 8.29 $\pm$ 2.57        & \textless{}3.92  & \textless{}122.01 & \textless{}3.83  \\
P4TU        & TISCO          & JP-I    & Stainless steel & 1.64      & mBq/kg & \textless{}2.19 & \textless{}2.21 & \textless{}30.42 & \textless{}6.57  & \textless{}7.65  & \textless{}3.80  & \textless{}105.29 & \textless{}5.71  \\
P4TV        & TISCO          & JP-I    & Stainless steel & 5.22      & mBq/kg & \textless{}1.80 & \textless{}2.31 & \textless{}21.05 & \textless{}5.81  & \textless{}5.07  & \textless{}15.16 & \textless{}34.59  & \textless{}3.63  \\
Threaded insert &  & JP-I & Stainless steel & - & mBq/pc & 126.97 $\pm$ 3.23 & \textless{}1.13 & \textless{}23.56 & \textless{}6.43 & 74.50 $\pm$ 3.61 & \textless{}80.45 & \textless{}47.95 & \textless{}1.89\\
Packing of   distillation tower &  & JP-I & Stainless steel & 0.58 & mBq/kg & 629.81 $\pm$ 11.15 & \textless{}5.80 & \textless{}114.68 & 39.22 $\pm$ 9.76 & 47.76 $\pm$ 5.89 & \textless{}17.97 & \textless{}177.53 & \textless{}7.30 \\
Copper bars & LUOYANG COPPER & JP-I    & Copper          & 1.71      & mBq/kg & \textless{}1.22 & \textless{}1.10 & \textless{}16.73 & \textless{}4.20  & \textless{}2.31  & \textless{}5.21  & \textless{}29.65  & \textless{}3.18  \\
Zr beans & NANJING YOUTIAN & JP-I & Zirconium & 1.00 & mBq/kg & \textless{}1.40 & \textless{}1.60 & \textless{}37.81 & \textless{}6.05 & \textless{}2.20 & \textless{}175.98 & \textless{}26.31 & \textless{}3.77 \\
\\ \hline

\textbf{PMT Components} \\
PMT R11410                            & Hamamatsu & JP-I    & -        & -         & mBq/pc & \textless{}2.34  & \textless{}1.85 & \textless{}22.34  & \textless{}7.82  & \textless{}3.06 & \textless{}28.29 & \textless{}56.48 & \textless{}3.99 \\
Faceplate of PMT R11410               & Hamamatsu & JP-I    & Quartz   & -         & mBq/pc & \textless{}0.53  & \textless{}0.42 & \textless{}7.08   & \textless{}1.23  & \textless{}0.40 & \textless{}2.52  & \textless{}3.69  & \textless{}0.66 \\
Pure alminum sheet used in PMT R11410 & Hamamatsu & JP-II    & Aluminum & 0.03     & mBq/kg & \textless{}19.64  & \textless{}66.97 & \textless{}1169.75   & \textless{}180.08  & \textless{}36.60 & \textless{}298.00  & \textless{}394.29  & \textless{}69.22 \\
Ceramic stem body   of PMT R11410 & Hamamatsu & JP-I & Ceramic & 0.496 & mBq/kg & \textless{}1.05 & \textless{}1.97 & 82.31 $\pm$ 13.26 & \textless{}7.07 & 7.30 $\pm$ 1.49 & \textless{}7.10 & 76.95 $\pm$ 18.12 & 15.58 $\pm$ 1.46 \\
Spring of PMT R11410 &  & JP-I & Stainless steel &  & mBq/pc & 1.24 $\pm$ 0.06 & \textless{}0.04 & \textless{}1.15 & \textless{}0.15 & \textless{}0.05 & \textless{}0.15 & \textless{}1.31 & \textless{}0.10\\
Base of PMT R11410 & - & JP-I & - & - & mBq/pc & \textless{}0.24 & \textless{}0.24 & \textless{}2.81 & \textless{}0.79 & \textless{}0.40 & \textless{}3.68 & \textless{}9.66 & 0.68 $\pm$ 0.21 \\
Base of PMT R11410 & - & JP-II & - & - & mBq/pc & \textless{}0.01 & \textless{}0.88 & \textless{}8.73 & \textless{}2.30 & \textless{}0.97 & \textless{}2.88 & 8.68 $\pm$ 2.82 & \textless{}1.65 \\
Spare base of PMT R11410 & - & JP-I & - & - & mBq/pc & \textless{}0.04 & \textless{}0.05 & 1.12 $\pm$ 0.34 & 0.24 $\pm$ 0.07 & 0.25 $\pm$ 0.05 & \textless{}1.22 & 7.47 $\pm$ 0.72 & 1.04 $\pm$ 0.05 \\
PCB of PMT R11410 (empty)             &           & JP-I    & Kapton   & -         & mBq/pc & \textless{}0.19  & \textless{}0.14 & \textless{}2.10   & \textless{}0.47  & \textless{}0.21 & \textless{}0.16  & \textless{}5.68  & \textless{}0.35 \\
Capacitor                             & KEMET     & JP-I    & -        & -         & mBq/pc & \textless{}0.01  & \textless{}0.01 & \textless{}0.26   & 0.35 $\pm$ 0.03        & 0.23 $\pm$ 0.02       & \textless{}0.01  & 1.24 $\pm$ 0.18        & 1.30 $\pm$ 0.03       \\
Capacitor                             & Johanson  & JP-I    & -        & -         & mBq/pc & \textless{}0.01  & \textless{}0.01 & \textless{}0.35   & 0.28 $\pm$ 0.04        & \textless{}0.01 & \textless{}0.06  & 7.88 $\pm$ 0.41        & 9.46 $\pm$ 0.09       \\
Capacitor                             & Vishay    & JP-I    & -        & -         & mBq/pc & \textless{}0.02  & \textless{}0.01 & \textless{}0.41   & 0.29 $\pm$ 0.03        & 0.14 $\pm$ 0.01       & \textless{}0.01  & 0.62 $\pm$ 0.13        & 0.68 $\pm$ 0.02       \\
Capacitor                             & AVX       & JP-I    & -        & -         & mBq/pc & \textless{}0.01  & \textless{}0.01 & \textless{}0.14   & \textless{}0.06  & 0.04 $\pm$ 0.01       & \textless{}0.01  & 0.47 $\pm$ 0.09        & 0.63 $\pm$ 0.02       \\
Capacitor                             & Yageo     & JP-I    & -        & -         & mBq/pc & \textless{}0.01  & \textless{}0.01 & \textless{}0.19   & \textless{}0.04  & \textless{}0.02 & \textless{}0.01  & \textless{}0.17  & 0.13 $\pm$ 0.01       \\
Capacitor                             & Knowles   & JP-I    & -        & -         & mBq/pc & \textless{}0.01  & \textless{}0.01 & \textless{}0.06   & \textless{}0.03  & \textless{}0.00 & \textless{}0.15  & \textless{}0.27  & 0.16 $\pm$ 0.01       \\
\hline

\end{tabular}
\end{adjustbox}
\end{sidewaystable*}

\begin{sidewaystable*}[tp]
\centering
\begin{adjustbox}{width=\textwidth,center}
\begin{tabular}{l l l l l l c c c c c c c c}
\multicolumn{14}{c}{\textbf{Table: continued}}\\
\toprule
Name        & Supplier       & Detector & Material        & Mass (kg) & Unit   & $^{60}$Co             & $^{137}$Cs            & $^{40}$K               & $^{232}$Th$_{e}$           & $^{232}$Th$_{l}$           & $^{235}$U             & $^{238}$U$_{e}$             & $^{238}$U$_{l}$            \\ \hline
Resistor (0 M${\Omega}$) &  & JP-I & - & - & mBq/pc & \textless{}1.53$\times 10^{-3}$ & \textless{}1.21$\times 10^{-3}$ & \textless{}1.88$\times 10^{-2}$ & \textless{}6.46$\times 10^{-3}$ & \textless{}2.96$\times 10^{-3}$ & \textless{}1.46$\times 10^{-3}$ & \textless{}1.17$\times 10^{-2}$ & \textless{}3.78$\times 10^{-3}$ \\
Resistor (5 M${\Omega}$) &  & JP-I & - & - & mBq/pc & \textless{}1.25$\times 10^{-3}$ & \textless{}9.90$\times 10^{-4}$ & 0.04 $\pm$ 0.01 & (9.69 $\pm$ 2.26)$\times 10^{-3}$ & (7.21 $\pm$ 1.17)$\times 10^{-3}$ & \textless{}1.13$\times 10^{-3}$ & \textless{}5.28$\times 10^{-2}$ & (1.01 $\pm$ 0.13)$\times 10^{-2}$ \\
Resistor (7.5 M${\Omega}$) &  & JP-I & - & - & mBq/pc & \textless{}2.24$\times 10^{-3}$ & \textless{}1.79$\times 10^{-3}$ & \textless{}2.21$\times 10^{-2}$ & \textless{}1.66$\times 10^{-2}$ & (1.09 $\pm$ 0.20)$\times 10^{-2}$ & \textless{}2.07$\times 10^{-3}$ & \textless{}6.96$\times 10^{-2}$ & (1.13 $\pm$ 0.20)$\times 10^{-3}$ \\
Resistor (10 M${\Omega}$) &  & JP-I & - & - & mBq/pc & \textless{}1.59$\times 10^{-3}$ & \textless{}1.03$\times 10^{-3}$ & 0.03 $\pm$ 0.01 & (1.29 $\pm$ 0.25)$\times 10^{-2}$ & (1.00 $\pm$ 0.13)$\times 10^{-2}$ & \textless{}1.14$\times 10^{-3}$ & \textless{}1.96$\times 10^{-2}$ & (1.34 $\pm$ 0.14)$\times 10^{-3}$ \\
Resistor (20 M${\Omega}$) &  & JP-I & - & - & mBq/pc & \textless{}1.49$\times 10^{-3}$ & \textless{}1.12$\times 10^{-3}$ & \textless{}1.75$\times 10^{-2}$ & \textless{}3.78$\times 10^{-3}$ & (3.74 $\pm$ 1.09)$\times 10^{-3}$ & \textless{}1.22$\times 10^{-3}$ & \textless{}1.02$\times 10^{-2}$ & \textless{}4.45$\times 10^{-3}$ \\
Resistor (100 k${\Omega}$) &  & JP-I & - & - & mBq/pc & \textless{}1.50$\times 10^{-3}$ & \textless{}1.25$\times 10^{-3}$ & (2.34 $\pm$ 0.77)$\times 10^{-2}$ & (1.16 $\pm$ 0.20)$\times 10^{-2}$ & (9.97 $\pm$ 1.07)$\times 10^{-3}$ & \textless{}1.77$\times 10^{-3}$ & \textless{}2.49$\times 10^{-2}$ & (1.66 $\pm$ 0.12)$\times 10^{-2}$\\
Socket of PMT R11410                  &           & JP-I    & -        & -         & mBq/pc & \textless{}0.01  & \textless{}0.01 & \textless{}0.12   & \textless{}0.05  & \textless{}0.01 & \textless{}0.01  & \textless{}0.29  & \textless{}0.01 \\
Elbow pin &  & JP-I & - & - & mBq/pc & \textless{}9.20$\times 10^{-4}$ & \textless{}1.26$\times 10^{-3}$ & \textless{}1.27$\times 10^{-3}$ & \textless{}9.90$\times 10^{-4}$ & \textless{}8.00$\times 10^{-4}$ & \textless{}6.70$\times 10^{-4}$ & \textless{}2.12$\times 10^{-3}$ & \textless{}7.70$\times 10^{-4}$ \\
Solder                                &           & JP-I    &          & 0.17     & mBq/kg & \textless{}10.71 & \textless{}8.28 & \textless{}174.58 & \textless{}31.07 & \textless{}7.67 & \textless{}9.83  & \textless{}73.35 & \textless{}9.65 \\
PMT R8520                             & Hamamatsu & JP-II    & -        & -         & mBq/pc & \textless{}0.61  & \textless{}1.11 & \textless{}23.09  & \textless{}2.78  & \textless{}0.66 & \textless{}5.57  & \textless{}10.92  & \textless{}1.50 \\
Base of PMT R8520                     & Hamamatsu & JP-II    & -        & -         & mBq/pc & \textless{}0.12  & \textless{}0.23 & \textless{}3.67   & \textless{}0.81  & \textless{}0.13 & \textless{}1.08  & \textless{}3.79  & 0.69 $\pm$ 0.14  \\    
Cable & KUNXINGSHENGDA & JP-I & Kapton & - & mBq/m & \textless{}0.01 & \textless{}0.01 & 1.25 $\pm$ 0.14 & \textless{}0.06 & \textless{}0.03 & \textless{}0.38 & \textless{}0.14 & \textless{}0.04 \\ \\
\hline
\textbf{PTFE} \\
PTFE \#1 & SANXIN & JP-I & PTFE & 0.89 & mBq/kg & \textless{}1.35 & \textless{}1.66 & \textless{}29.28 & \textless{}4.24 & \textless{}2.79 & \textless{}3.61 & \textless{}50.32 & \textless{}2.37 \\
PTFE \#2 & SANXIN & JP-I & PTFE & 1.79 & mBq/kg & \textless{}1.32 & \textless{}1.35 & \textless{}26.64 & \textless{}5.48 & 5.15 $\pm$ 1.41 & \textless{}4.12 & \textless{}13.91 & \textless{}3.20 \\
PTFE \#3 & SANXIN & JP-I & PTFE & 1.23 & mBq/kg & \textless{}2.76 & \textless{}2.77 & \textless{}25.28 & \textless{}10.50 & \textless{}9.01 & \textless{}19.73 & \textless{}128.38 & \textless{}4.36 \\
PTFE \#4 & SANXIN & JP-I & PTFE & 0.61 & mBq/kg & \textless{}2.95 & \textless{}2.86 & \textless{}66.14 & \textless{}7.18 & \textless{}2.33 & \textless{}6.61 & \textless{}27.00 & \textless{}5.08 \\
PMT R8520 holder \#1 & DONGYUE & JP-I & PTFE & 0.39 & mBq/kg & \textless{}3.46 & \textless{}3.99 & \textless{}116.51 & \textless{}13.60 & \textless{}8.32 & \textless{}5.23 & \textless{}37.71 & \textless{}4.58 \\
PMT R8520 holder \#2 & DONGYUE & JP-I & PTFE & 1.28 & mBq/kg & \textless{}1.95 & \textless{}1.97 & \textless{}49.05 & \textless{}5.99 & \textless{}3.53 & \textless{}2.90 & \textless{}22.44 & \textless{}2.63 \\

\bottomrule

\end{tabular}%
\end{adjustbox}
\end{sidewaystable*}

\begin{sidewaystable*}[h]
\scriptsize
\centering
\caption{Radioactivity input in the background estimation}
\label{table:input}
\begin{adjustbox}{width=\textwidth,center}
\begin{tabular}{lccccccccc}
\toprule
 & unit & $^{60}$Co & $^{137}$Cs & $^{40}$K & $^{232}$Th$_{e}$ & $^{232}$Th$_{l}$ & $^{235}$U & $^{238}$U$_{e}$ & $^{238}$U$_{l}$ \\
 \hline
PMT R11410 & mBq/pc & 1.16 $\pm$ 0.72 & 0.52 $\pm$ 0.81 & 8.37 $\pm$ 8.47 & 4.33 $\pm$ 2.16 & 1.50 $\pm$ 0.96 & 13.13 $\pm$ 8.53 & 26.29 $\pm$ 16.90 & 2.05 $\pm$ 1.18 \\
Window of PMT R11410 & mBq/pc & 0.00 $\pm$ 0.32 & 0.05 $\pm$ 0.22 & 1.78 $\pm$ 3.22 & 0.00 $\pm$ 0.75 & 0.08 $\pm$ 0.20 & 0.00 $\pm$ 2.24 & 0.13 $\pm$ 0.32 & 0.00 $\pm$ 2.24 \\
Stem of PMT R11410 & mBq/kg & 0.26 $\pm$ 0.48 & 0.97 $\pm$ 0.61 & 82.31 $\pm$ 13.26 & 4.07 $\pm$ 1.82 & 7.30 $\pm$ 1.49 & 3.74 $\pm$ 2.04 & 76.95 $\pm$ 18.12 & 15.58 $\pm$ 1.46 \\
Base of PMT R11410 & mBq/pc & 0.03 $\pm$ 0.05 & 0.00 $\pm$ 0.28 & 0.76 $\pm$ 2.41 & 0.28 $\pm$ 0.62 & 0.28 $\pm$ 0.18 & 0.46 $\pm$ 1.22 & 6.97 $\pm$ 1.94 & 0.84 $\pm$ 0.22 \\
Spring of PMT R11410 & mBq/pc & 0.04 $\pm$ 0.01 & 0.00 $\pm$ 0.01 & 0.00 $\pm$ 0.03 & 0.01 $\pm$ 0.01 & 0.02 $\pm$ 0.01 & 0.00 $\pm$ 0.01 & 0.00 $\pm$ 0.03 & 0.01 $\pm$ 0.01 \\
PMT R8520 & mBq/pc & 0.28 $\pm$ 0.20 & 0.11 $\pm$ 0.61 & 9.70 $\pm$ 8.14 & 0.23 $\pm$ 1.55 & 0.00 $\pm$ 0.40 & 0.59 $\pm$ 3.03 & 3.30 $\pm$ 4.63 & 0.51 $\pm$ 0.60 \\
Base of PMT R8520 & mBq/pc & 0.00 $\pm$ 0.07 & 0.02 $\pm$ 0.13 & 0.00 $\pm$ 2.36 & 0.25 $\pm$ 0.31 & 0.00 $\pm$ 0.07 & 0.00 $\pm$ 0.57 & 1.89 $\pm$ 0.95 & 0.64 $\pm$ 0.13 \\
IV barrel and IV dome & mBq/kg & 1.07 $\pm$ 1.26 & 0.17 $\pm$ 1.28 & 8.89 $\pm$ 15.51 & 0.36 $\pm$ 4.42 & 0.72 $\pm$ 1.82 & 0.32 $\pm$ 2.67 & 30.23 $\pm$ 41.16 & 1.17 $\pm$ 2.04 \\
OV barrel & mBq/kg & 0.00 $\pm$ 1.27 & 0.00 $\pm$ 1.16 & 26.13 $\pm$ 13.42 & 0.20 $\pm$ 2.86 & 1.34 $\pm$ 1.83 & 5.26 $\pm$ 2.71 & 40.87 $\pm$ 22.77 & 0.53 $\pm$ 1.44 \\
OV domes and electrodes & mBq/kg & 0.51 $\pm$ 1.04 & 0.00 $\pm$ 0.99 & 3.00 $\pm$ 9.84 & 2.48 $\pm$ 2.57 & 3.17 $\pm$ 1.72 & 2.78 $\pm$ 2.42 & 40.84 $\pm$ 24.03 & 1.51 $\pm$ 1.33 \\
Flanges and bolt screws & mBq/kg & 0.00 $\pm$ 1.15 & 0.00 $\pm$ 1.09 & 25.51 $\pm$ 12.52 & 3.27 $\pm$ 2.77 & 2.60 $\pm$ 1.82 & 2.81 $\pm$ 1.90 & 0.00 $\pm$ 15.81 & 0.48 $\pm$ 1.33 \\
Threaded insert & mBq/pc & 43.93 $\pm$ 1.15 & 0.25 $\pm$ 0.20 & 1.74 $\pm$ 2.83 & 1.27 $\pm$ 0.83 & 19.45 $\pm$ 1.01 & 0.99 $\pm$ 0.72 & 7.94 $\pm$ 5.73 & 0.00 $\pm$ 0.31 \\
Shaping rings and copper plates & mBq/kg &  &  &  & (5.16 $\pm$ 1.38)$\times 10^{-3}$ &  &  & (5.59 $\pm$ 0.15)$\times 10^{-2}$ &  \\
PTFE holders and reflectors & mBq/kg &  &  & (4.65 $\pm$ 0.19)$\times 10^{-5}$ & 0.04 $\pm$ 0.01 &  &  & \textless{}0.01 & \\
\bottomrule
\end{tabular}%
\end{adjustbox}
\end{sidewaystable*}

\clearpage

\end{document}